\documentclass[
showpacs,
floatfix,
aps,  
prl,
twocolumn,
superscriptaddress,
amssymb,
tightenlines,
groupaddress,
]{revtex4-1}

\usepackage{times}
\usepackage{wasysym}
\usepackage{amssymb}
\usepackage{latexsym}
\usepackage[dvips]{graphicx}
\usepackage{graphicx}
\usepackage{dcolumn}
\usepackage{amsfonts}
\usepackage{bm}
\usepackage{epsfig}
\usepackage{xcolor}
\usepackage{soul}

\newcommand{\be}{\begin{equation}}
\newcommand{\ee}{\end{equation}}
\newcommand{\bea}{\begin{eqnarray}}
\newcommand{\eea}{\end{eqnarray}}

\def\nn{\nonumber}

\def\m{m^\star}

\def\oc{\omega_{\mbox{\scriptsize {c}}}}

\def\rc{R_{\mbox{\scriptsize {c}}}}

\def\tpi{\tau_{\pi}}

\def\tq{\tau_{\rm q}}
\def\ttr{\tau}

\def\tin{\tau_{\rm in}}

\def\lb{l_B}

\newcommand{\req}[1]{Eq.\,(\ref{#1})}
\newcommand{\rEq}[1]{Equation\,(\ref{#1})}

\newcommand{\rfig}[1]{Fig.\,\ref{#1}}

\newcommand{\rref}[1]{Ref.\,\onlinecite{#1}}

\def\edc{\epsilon}

\def\ne{n_s}

\def\B{{\bf B}}
\def\E{{\bf E}}

\def\k{{\bf k}}
\def\q{{\bf q}}
\def\v{{\bf v}}

\begin{document}
\title{Effect of Berry phase on nonlinear response of two-dimensional fermions}

\author{O. E. Raichev}
\email[Corresponding author: ]{raichev@isp.kiev.ua}
\affiliation{Institute of Semiconductor Physics, NAS of Ukraine, Prospekt Nauki 41, 03028 Kyiv, Ukraine}
\author{M. A. Zudov}
%\email[Corresponding author: ]{zudov001@umn.edu}
\affiliation{School of Physics and Astronomy, University of Minnesota, Minneapolis, Minnesota 55455, USA}
\received{\today}

\begin{abstract}
We develop a theory of nonlinear response to an electric field of two-dimensional (2D) fermions 
with topologically non-trivial wave functions characterized by the Berry phase $\Phi_n = n \pi, n = 1,2,...$\,. 
In particular, we find that owing to suppression of backscattering at odd $n$, Hall field-induced resistance oscillations, which stem from elastic electron transitions between Hall field-tilted Landau levels, are qualitatively distinct from those at even $n$: their amplitude decays with the electric field and their extrema are phase-shifted by a quarter cycle.
The theory unifies the cases of graphene ($n = 1$) and graphite bilayer ($n = 2$) with the case of conventional 2D electron gas ($n = 0$) and suggests a new method to probe backscattering in topological 2D systems.
\end{abstract}

\maketitle

The topological property of two-dimensional (2D) massless Dirac fermions, expressed in terms of the Berry phase $\Phi_1=\pi$, is responsible for peculiar Landau quantization manifesting itself in the phase-shifted Shubnikov-de Haas oscillations (SdHO) and unconventional quantum Hall effect \citep{novoselov:2005,zhang:2005,gusynin:2005a,gusynin:2005b}. 
This makes the Dirac fermions in graphene fundamentally distinct from both the conventional 2D electron gas (2DEG) in quantum wells ($\Phi_0=0$) and topologically non-trivial fermions in graphite bilayers ($\Phi_2=2\pi$) \citep{novoselov:2006,mccann:2006}. 
Another immediate consequence of the topological nature of wave functions of massless Dirac fermions is the absence of elastic backscattering off scalar potentials, which has various manifestations in electronic 
properties. 
In particular, it lies at the origin of Klein tunneling \citep{beenakker:2008} and its implications for resistivity of graphene-based $n$-$p$-$n$ junctions \citep{katsnelson:2006,cheianov:2006b,li:2018}. 
Furthermore, it removes a sharp cusp of static polarizability of degenerate carriers at doubled Fermi wavenumber (characteristic for conventional 2DEG and graphite bilayer) \citep{kotov:2012}, leading to enhanced spatial decay of Friedel oscillations \citep{cheianov:2006a}. 

Another well known effect which crucially depends on backscattering is Hall field-induced resistance oscillations (HIRO) \citep{yang:2002,zhang:2007a,vavilov:2007} which emerge in differential resistivity $r$ of a 2DEG subjected to elevated current density $j$ and perpendicular magnetic field $B$.
These oscillations appear due to the property of enhanced phase space for elastic transitions 
in 2DEG in the vicinity of backscattering.
In the presence of classically strong $B$, the backscattering means a spatial shift of the cyclotron orbit guiding center twice the cyclotron radius $\rc$, so the transition rate increases each instant when $2\rc$ equals integer multiple of spatial separation between Landau levels (LLs) tilted by the electric field $E$. 
Therefore, $r$ oscillates with the dimensionless parameter $\edc= 2 \rc |e|E /\hbar \oc$,
where $E \simeq E_H$, $E_H$ is the Hall electric field, and $\oc$ is the cyclotron frequency. As a result, 
HIRO are periodic in $1/B$ with the frequency ($\edc \equiv B_1/B$)
\be
B_1 = \frac {8 \pi \hbar c}{{\rm g} e^2 v_F}j\,,
%2
\label{eq.b1}
\ee
where $v_F$ is the group velocity at the Fermi level and ${\rm g}$ is the degree of band degeneracy.

To date, HIRO have been observed only in topologically trivial ($n=0$) 2D systems based on GaAs/ AlGaAs \citep{yang:2002,zhang:2007a}, Ge/SiGe \citep{shi:2014b} and ZnO/MgZnO \citep{shi:2017b} heterostructures. 
In overlapping LLs, they are described by quantum correction to the differential resistivity $\delta r$ \citep{vavilov:2007}
\begin{equation}
\frac {\delta r} {\rho_0} = 2 \lambda^2 \mathcal{H}_0(\edc)\,,~~\mathcal{H}_0(\edc) = \frac{8}{\pi}\frac{\ttr}{\tpi} \cos 2\pi\edc\,,~~ \pi\edc \gg 1\,,
%1
\label{eq.hiro}
\end{equation}
where $\rho_0$ is the resistivity at $B=0$, $\ttr$ is the transport scattering time, $\tpi$ is the backscattering time, $\lambda = \exp(-\pi/\oc\tq)$ is the Dingle factor, and $\tq$ is the quantum lifetime.
As a result, at $n = 0$ HIRO are described by $\mathcal{H}_0(\edc)$ whose maxima occur at $\edc= m = 1, 2, 3, ...$ and whose amplitude is independent of $\edc$.

In this paper we demonstrate that the nonlinear response of topologically non-trivial ($n \ge 1$) 
2D systems crucially depends on the parity of $n$. For {\em even} $n$, backscattering is not 
suppressed and the behavior is qualitatively the same as in conventional 2DEGs.
However, suppression of backscattering in massless Dirac systems, as well as in any other system with 
{\em odd} $n$, leads to a decay of the HIRO amplitude (as $1/\edc$) and produces a characteristic 
quarter-cycle phase shift of the oscillations towards larger $\edc$. In particular, for Dirac fermions 
($n=1$) we obtain [see the definition of $\tpi$ for this case after \req{eq.hca1}]
\begin{equation}
\frac {\delta r} {\rho_0} = 2 \lambda^2 \mathcal{H}_1(\edc)\,,~~\mathcal{H}_1(\edc) = \frac{16}{\pi^2}\frac{\ttr}{\tpi} \frac {\sin 2\pi\edc}{\edc}\,,~~ \pi\edc \gg 1\,,
%4
\label{eq.hiro1}
\end{equation}
which describes HIRO with the frequency given by \req{eq.b1} and the maxima near $\edc = m + 1/4$. 
We also find that the HIRO amplitude increases with $n$ for odd $n$. Our conclusion that the absence 
of backscattering leads to a phase shift of nonlinear magnetoresistance oscillations can become a basis 
for a new method to probe backscattering in topological 2D systems.

Our theory of nonlinear magnetotransport is developed for the regime of large occupation factors 
(high LLs), classically strong magnetic fields, and overlapping LLs, which is relevant for 
observation of HIRO \citep{dmitriev:2012}. 
We consider spin-degenerate 2D systems described by the Hamiltonian
\be
{\hat H} = \varepsilon(k) \left( \begin{array}{cc} 0 & e^{-i n \varphi} \\ 
e^{i n \varphi} & 0 \end{array} \right)\,,
%3
\label{eq.Ham}
\ee
where $\varphi$ is the angle of the wave vector $\k$, the energy spectrum $\pm \varepsilon(k)$ is 
isotropic, and the winding number $n$ gives the Berry phase $\Phi_n=n \pi$.
The Hamiltonian ${\hat H}$ describes fermions in graphene ($K$ valley) at $n=1$ 
[$\varepsilon(k)=\hbar v_F k$ with constant Fermi velocity $v_F \simeq 10^8$ cm/s] 
and in graphite bilayer at $n=2$ [$\varepsilon(k)=(\hbar k)^2/2\m$ with constant 
effective mass $\m \simeq 0.037$ of free electron mass]. For these particular systems, 
$n$ can be viewed as the degree of chirality in the carbon sublattice space \citep{mccann:2006}. 
Though ${\hat H}$ formally produces a two-band spectrum, we consider only intraband 
excitations and the topologically trivial case ($n=0$) can be equally applied to 2DEG 
in quantum wells. 

To derive expressions for the resistivity, we use \req{eq.Ham} and consider elastic scattering 
of fermions by impurities. Adopting the methods developed for 2DEG with parabolic spectrum \citep{vavilov:2004,dmitriev:2012}, in particular, using the reference frame moving with the 
drift velocity \citep{vavilov:2004,vavilov:2007}, we can write the steady-state Boltzmann 
equation for the distribution function $f_{\varepsilon,\varphi}$ of a 2DEG placed in perpendicular 
magnetic field $\B = (0,0, B)$ and in-plane electric field $\E = (E_x,E_y,0)$ in the following form:
\begin{eqnarray}
\oc \frac{\partial f_{\varepsilon,\varphi}}{\partial \varphi} &=& 
\frac{2 \pi}{\hbar}
\int\limits_0^{2 \pi} \frac{d \varphi'}{2 \pi} \int d \varepsilon' \delta(\varepsilon'-\varepsilon-\gamma) \nn \\
&\times& w_{\k \k'} F^{(n)}_{\k \k'} \nu_{\varepsilon'} (f_{\varepsilon',\varphi'}-f_{\varepsilon,\varphi})\,.
%5
\label{eq.col}
\end{eqnarray}
The right-hand side is the collision integral describing elastic scattering of electrons by impurities within a single valley. Here, $\gamma=\hbar \v_D \cdot \q$ is the work of the electric force $|e|\E$ during a scattering event, $\v_D=c[\E \times \B]/B^2$ is the drift velocity, $\q=\k-\k'$ is the wave vector transmitted in scattering, $\nu_{\varepsilon}$ is the density of states per spin and valley, and $w_{\k \k'}$ is the Fourier transform 
of the correlation function of the impurity potential. 
It is assumed below that $\gamma \ll \varepsilon$ so that $|\k' | \simeq |\k| \equiv k_{\varepsilon}$.
The cyclotron frequency $\oc = |e| B/\m c$ is determined 
by the effective (cyclotron) mass $\m=\hbar k_{\varepsilon}/v_F$, which, in general, is energy-dependent 
(for graphene, $k_{\varepsilon}=\varepsilon/\hbar v_F$). The function
\begin{equation}
F^{(n)}_{\k \k'}=F^{(n)}_{\theta} =(1+\cos n \theta)/2
%6
\label{eq.ff}
\end{equation}
is the squared overlap integral of columnar eigenstates of ${\hat H}$ with wave vectors $\k$ and 
$\k'$, and $\theta=\varphi-\varphi'$ is the scattering angle. 
It is seen directly that backscattering ($\theta=\pi$) is suppressed 
for odd $n$ but not for even $n$. 

For overlapping LLs, the oscillatory density of states at $\varepsilon \gg \hbar \oc$ follows from the Bohr-Sommerfeld quantization rule corrected by the Berry phase \citep{xiao:2010}: 
\begin{equation}
\nu_{\varepsilon} \simeq {\overline \nu}_{\varepsilon} \left[ 1 -  2 \lambda_{\varepsilon} 
\cos \left(\pi k^2_{\varepsilon} \lb^2 + \Phi_n \right) \right]\,,
%6
\label{eq.dos}
\end{equation}
where ${\overline \nu}_{\varepsilon}$ is the density of states at $B = 0$ (for graphene, ${\overline \nu}_{\varepsilon}=\varepsilon/2 \pi \hbar^2 v_F^2$), $\lb = \sqrt{\hbar c/|e|B}$ is the magnetic length, 
and $\lambda_{\varepsilon}$ is the Dingle factor at energy $\varepsilon$. The quantity $k^2_{\varepsilon} 
\lb^2/2$ is the number of magnetic flux quanta inside the cyclotron orbit of electron with 
energy $\varepsilon$. \rEq{eq.dos} can also be
derived from the self-consistent Born approximation (for graphene, see, e.g., \rref{briskot:2013}). 
The density of states has maxima at the Landau quantization energies, $\varepsilon_N= 
\hbar v_F \lb^{-1} \sqrt{2 N}$ ($N = 0, 1, 2, ...$\,) for graphene and $\varepsilon_N= \hbar 
\oc(N-1/2)$ for graphite bilayer. In the latter case, $\varepsilon_N$ approximates the exact spectrum 
$\varepsilon_N= \hbar \oc \sqrt{N(N-1)}$ \citep{mccann:2006} at $N \gg 1$.

The length of $\q$ and its angle $\varphi_q$ can be expressed as
\begin{equation}
q=2k_{\varepsilon}\sin \frac{\theta}{2} \equiv q_{\varepsilon,\theta}\,,~~
\varphi_q=\phi+\pi/2\,,~~ \phi \equiv \frac{\varphi+\varphi'}{2}\,,
%4
\label{eq.q}
\end{equation}
and the work $\gamma$ can be conveniently rewritten as 
\begin{equation}
\gamma = -2 \hbar k_{\varepsilon} v_D \cos \phi' \sin \frac{\theta}{2}\,,~~ \phi'\equiv \phi-\chi\,, 
%5
\label{eq.gamma}
\end{equation}
where $\chi$ is the angle of the electric field $\E$. The assumed strong inequality 
$\gamma \ll \varepsilon$ always holds at $v_D \ll v_F$.

\rEq{eq.col} is easily solved in the regime of classically strong magnetic fields ($v_F \tau \gg 
k_{\varepsilon} l_B^2$), when one can replace the distribution 
functions under the integral by an isotropic distribution $f_{\varepsilon}$. Substituting such a 
solution into the expression for the current density,
\begin{eqnarray}
{\bf j}={\bf j}^{(d)} + e n_s {\bf v}_D,~
{\bf j}^{(d)}={\rm g} e \int d \varepsilon \nu_{\varepsilon} \int\limits_0^{2 \pi} \frac{d \varphi}{2 \pi} 
{\bf v}_F f_{\varepsilon,\varphi}, 
\label{eq.j}
\end{eqnarray}
where $\ne$ is the carrier density, and taking into 
account that the impurity potential correlator depends only on the absolute value of $\q$, $w_{\k \k'}=w(q)$, we obtain the dissipative conductivity (${\bf j}^{(d)} = \sigma_d \E$): 
\begin{eqnarray}
\sigma_d&=&\frac{{\rm g} \pi \hbar c^2}{B^2}
\int d \varepsilon k_{\varepsilon}^2 \int\limits_0^{2 \pi} \frac{d \phi'}{2 \pi} 
\int\limits_0^{2 \pi} \frac{d \theta}{2 \pi}  w(q_{\varepsilon,\theta}) F^{(n)}_{\theta} \gamma^{-1}
\nn\\
&\times&  (1+ \cos 2 \phi') (1-\cos \theta) \nu_{\varepsilon} \nu_{\varepsilon+\gamma} 
(f_{\varepsilon} -f_{\varepsilon+\gamma})\,.
%8
\label{eq.sigma}
\end{eqnarray}
The isotropic part of the distribution function standing in \req{eq.sigma} can be represented as a sum of quasiequilibrium Fermi distribution $f^{(0)}_{\varepsilon}=\{ \exp[(\varepsilon-\varepsilon_F) /k_B T_e]+1\}^{-1}$, where $T_e$ is the temperature of carriers and $k_B$ is the Boltzmann constant, and a small non-equilibrium contribution $\delta f_{\varepsilon}$ caused by the field-induced redistribution of electrons in the energy domain in the presence of Landau quantization \citep{dmitriev:2005,vavilov:2007}. 
The function $\delta f_{\varepsilon}$ shows rapid oscillations similar to those in $\nu_\varepsilon$, \req{eq.dos}, and can be found from \req{eq.col} averaged over $\varphi$. 
To describe relaxation of the isotropic distribution, the inelastic relaxation term $-\delta f_{\varepsilon}/\tin$ ($\tin$ is the inelastic relaxation time), approximating the linearized collision integral for electron-electron scattering, should be added to the right-hand side of \req{eq.col}. 
To the first order in $\lambda_\varepsilon$,
\be
\delta f_{\varepsilon} \simeq -\frac{2 \lambda_{\varepsilon} \hbar k_F v_D
[\partial \eta_n(\zeta)/\partial \zeta] }{\tin^{-1}+\tq^{-1}-\eta_n(\zeta)} \frac{\partial f^{(0)}_{\varepsilon}}{\partial \varepsilon} \sin(\pi k_{\varepsilon}^2 \lb^2+\Phi_n)\,,
\label{eq.deltaf}
\ee
where $k_F=\sqrt{4 \pi\ne/{\rm g}}$ is the Fermi wavenumber. 
The quantum and the transport scattering rates are given by
\begin{equation}
\frac{1}{\tq}=\int\limits_0^{\pi} \frac{d \theta}{\pi} \frac{1}{\tau_n(\theta)}\,,~~
\frac{1}{\ttr}=\int\limits_0^{\pi} \frac{d \theta}{\pi} \frac{1-\cos \theta}{\tau_n(\theta)}\,, 
\label{eq.tau}
\end{equation}
respectively, with
\begin{equation}
\frac{1}{\tau_n(\theta)}= \frac{2 \pi}{\hbar} {\overline \nu}_{\varepsilon_F} w(q_{\varepsilon_F,\theta}) F^{(n)}_{\theta}\,. 
%11
\label{eq.tau1}
\end{equation}
Next,
\begin{equation}
\eta_n(\zeta)= \int\limits_0^{\pi} \frac{d \theta}{\pi} \frac{1}{\tau_n(\theta)} 
J_0 \left(2 \zeta \sin \frac{\theta}{2} \right)\,, 
%11
\label{eq.eta}
\end{equation}
where $J_\alpha$ denotes a Bessel function of the first kind, and 
\begin{equation}
\zeta = \pi \edc\,,~~ \edc = 2 k_F^2 \lb^2 \frac {v_D}{v_F} \simeq 
\frac{8 \pi \hbar c}{{\rm g} e^2 v_F} \frac{j}{B}\,.
%13
\label{eq.edc}
\end{equation}
Assuming degenerate carriers, we have set $\varepsilon=\varepsilon_F$, $k_{\varepsilon}=k_F$ in all quantities whose energy dependence is weak. 

Substituting $f_{\varepsilon}=f^{(0)}_{\varepsilon}+ \delta f_{\varepsilon}$ into \req{eq.sigma} 
and calculating the integrals over $\varepsilon$ and $\phi'$ analytically, we represent  
the result as an expansion in powers of the Dingle factor, 
\begin{equation}
\sigma_d = \sigma_d^{(0)}+\sigma_d^{(1)}+\sigma_d^{(2)}\,
%17
\label{eq.totalsigma}
\end{equation} 
where $\sigma_d^{(0)}=\ne \hbar k_F c^2/v_F \ttr B^2$ is the Drude conductivity and
\bea
\sigma_d^{(1)} &=&-\sigma_d^{(0)} 4 \lambda {\cal D}_T \cos (\pi k_F^2 \lb^2+\Phi_n) s_n(\zeta)
%18
\label{eq.SdH}
\eea
is the term describing SdHO, where 
${\cal D}_T=X_T/\sinh X_T$, with $X_T= 2 \pi^2 k_B T_e k_F \lb^2/\hbar v_F$. 
The influence of the electric field on SdHO is described by
\be
s_n(\zeta) = - \frac{\ttr}{\zeta} \frac{\partial \eta_n(\zeta)}{\partial \zeta}\,, 
%19
\label{eq.s1}
\ee
as $\delta f_{\varepsilon}$ does not contribute to $\sigma_d^{(1)}$ \citep{dmitriev:2011}.
SdHO are strongly suppressed by temperature at $X_T \gg 1$, owing to rapid oscillations of 
$\nu_{\varepsilon}$. In the last term, $\sigma_d^{(2)}$, we retain only the contributions 
that survive at $X_T \gg 1$. We find
\bea
&\sigma_d^{(2)}& =\sigma_d^{(0)} 2 \lambda^2 [ h_n(\zeta)+ g_n(\zeta)]\,,
%20
\label{eq.sigma2}
\eea
where the parts  
\bea
h_n(\zeta) & = & -\ttr \frac{\partial^2 \eta_n(\zeta)}{\partial \zeta^2}
%21
\label{eq.h1}
\eea
and 
\be
g_n(\zeta) = - \frac{2 \ttr [\partial \eta_n(\zeta)/\partial \zeta]^2 }{
\tin^{-1}+\tq^{-1}-\eta_n(\zeta)}\,,
%22
\label{eq.g}
\ee
respectively, come from substitution of 
$f^{(0)}_{\varepsilon}$ and $\delta f_{\varepsilon}$ into \req{eq.sigma} and are often 
referred to as the displacement and the inelastic contributions. 

For $n = 0$, the results given by \req{eq.eta}$-$\req{eq.g} reproduce those obtained previously for the 2DEG with parabolic spectrum \citep{vavilov:2007,dmitriev:2011}.
However, they now carry important topological distinction between even and odd $n$ because the 
field-dependent quantities $\eta$, $s$, $h$, and $g$ depend on the Berry phase through the overlap factor $F^{(n)}_{\theta}$ in \req{eq.tau1}. 
To demonstrate this distinction, we consider the case of sharp ($\delta$-correlated) scattering potential, 
for which $w(q_{\varepsilon_F,\theta})$ is constant and $\ttr/\tq=1+\delta_{n,1}$,
and the integral in \req{eq.eta} can be evaluated analytically 
[below, $J_\alpha \equiv J_\alpha(\zeta)$]:
\be
\eta_n(\zeta) = (J_0^2+J_n^2)/\ttr \xi_n\,,~ \xi_{n}=1+\delta_{n,0}-\delta_{n,1}/2\,. 
%23
\label{eq.int}
\ee
We further find
\begin{equation}
s_n(\zeta)=[ 2J_0J_1+ (J_{n+1}-J_{n-1}) J_n ]/\zeta \xi_n\,,
%24
\label{eq.s2}
\end{equation}
\begin{eqnarray}
h_n(\zeta)&=&\left [ J^2_0-2J^2_1-J_0J_2 - (J_{n+1}-J_{n-1})^2/2\right. \nn \\
&-&J_{n}(J_{n-2} - 2J_n + J_{n+2})/2 \left. \right ]/\xi_n\,,~~
%25
\label{eq.h2}
\end{eqnarray}
and 
\be
g_n(\zeta) = -\frac {2 [\zeta s_n(\zeta)]^2}{\ttr/\tin + 1 + \delta_{n,1} - (J_0^2+J_n^2)/\xi_n}\,.
%26
\label{eq.g2}
\ee

In the weak field limit, $\zeta \ll 1$, the above expressions describe quadratic in $E$ corrections 
to the conductivity. In particular, $s_n=1-\mu_n \zeta^2$ and $h_n=1-3\mu_n \zeta^2$, where $\mu_0=3/8$, 
$\mu_1=1/4$, $\mu_2=7/16$, and $\mu_n=\mu_0$ for $n>2$. Remarkably, the function $g_n$ in this limit does 
not depend on $n$: $g_n=-2 \zeta^2/(\ttr/\tin+\zeta^2/2)$. 
If the inelastic relaxation is slow, $\ttr/\tin \ll 1$, this function gives the main contribution to the 
weak-field nonlinear response in $\sigma^{(2)}$. 

In the strong field limit, $\zeta \gg 1$, \req{eq.s2}$-$\req{eq.g2} describe field-induced 
oscillations of conductivity caused by transitions of carriers between different LLs. The 
result depends on the parity of $n$ in an essential way because the influence of backscattering 
on the conductivity becomes important. For even $n$,   
\bea
s_n(\zeta)&=&-\frac{4 \cos 2 \zeta}{\pi \zeta^2 (1+\delta_{n,0})}\,,~
h_n(\zeta)=\frac{8 \sin 2 \zeta}{\pi \zeta (1+\delta_{n,0})}, \nn \\
g_n(\zeta)&=&-\frac{16(1+ \cos 4 \zeta)}{\pi^2 \zeta^2 (\ttr/\tin+1)(1+\delta_{n,0})^2}\,,
%27
\label{eq.strongEeven}
\eea
whereas for odd $n$,
\bea
s_n(\zeta)&=&\frac{2 (1-n^2 \sin 2 \zeta)}{\pi \zeta^3 (1-\delta_{n,1}/2)}\,,~
h_n(\zeta)=-\frac{4 n^2 \cos 2 \zeta}{\pi \zeta^2 (1-\delta_{n,1}/2)}, \nn \\
g_n(\zeta)&=&-\frac{8(1+\delta_{n,1})^2(1-n^2 \sin 2 \zeta)^2}{\pi^2 \zeta^4 (\ttr/\tin+1+\delta_{n,1})}\,.
%28
\label{eq.strongEodd}
\eea
For odd $n$, all these quantities decrease with $\zeta$ faster than for even $n$ and the 
oscillations gain in amplitude with increasing $n$. Regardless of the parity of $n$, the main 
contribution to the strong-field response comes from $h_n$. This contribution describes HIRO, 
whose behavior in conventional 2DEG and in graphite bilayer ($n=2$) is predicted to be qualitatively 
the same, while in graphene ($n=1$) their behavior is essentially different.       

\begin{figure}[t]
\includegraphics{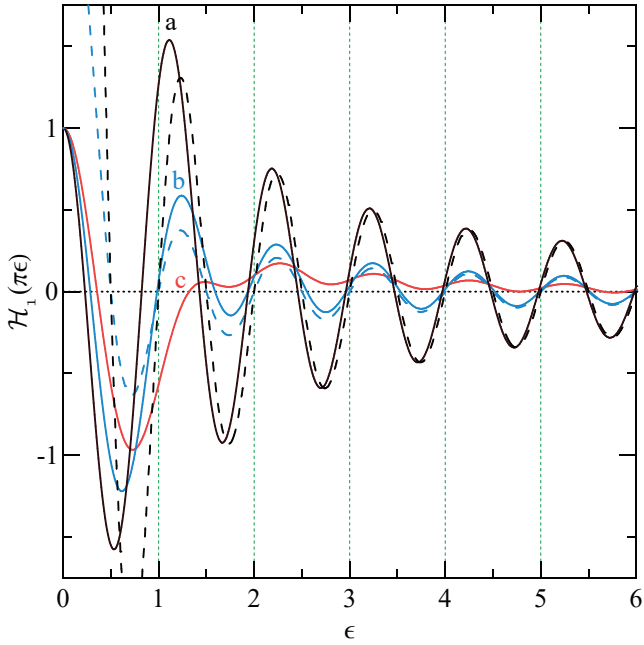}
\vspace{-0.1 in}
\caption{(Color online)
Function $\mathcal{H}_1(\pi\edc)$ for sharp scattering potential (solid curve a), and for smooth scattering potential modeled by $w(q) \propto e^{-\alpha q}$, where $\alpha$ is chosen to provide $\ttr/\tq=5$ (solid curve b) and $\ttr/\tq=10$ (solid curve c). 
The dashed lines represent the limit $\pi \edc \gg 1$ for curves a and b.}
\vspace{-0.15 in}
\label{fig1}
\end{figure}

We next focus on Dirac fermions, $n=1$. 
Using \req{eq.totalsigma}, we obtain the longitudinal resistivity,
\begin{eqnarray}
\frac {\rho} {\rho_0} &=& 1 +  4 \lambda {\cal D}_T \cos (\pi k_F^2 \lb^2) s_1(\zeta) \nn \\
&+& 2 \lambda^2 [h_1(\zeta)+g_1(\zeta)]\,,~~ \rho_0= \frac{\hbar k_F}{e^2 \ne v_F \ttr}\,.
%29
\label{eq.rho}
\end{eqnarray}
In experiments, one usually measures the differential resistivity 
$r \equiv \partial (j \rho)/ \partial j$. Then, $r=\rho_0 + \delta r$, where
\begin{eqnarray}
\frac{\delta r}{\rho_0} &=&
4 \lambda \cos (\pi k_F^2 \lb^2) \left\{  \left[ \partial (j {\cal D}_T)/\partial j \right] s_1(\zeta) 
+ {\cal D}_T \mathcal{S}_1(\zeta) \right\} \nn \\
&+& 2 \lambda^2 \left[ \mathcal{H}_1 (\zeta) + \mathcal{G}_1 (\zeta) \right ],~~
%30
\label{eq.r}
\end{eqnarray}
with $\mathcal{S}_1 (\zeta) \equiv \partial [\zeta s_1(\zeta)]/\partial \zeta$, $\mathcal{H}_1(\zeta) \equiv \partial 
[\zeta h_1(\zeta)]/ \partial \zeta$, $\mathcal{G}_1(\zeta) \equiv \partial [\zeta g_1(\zeta)]/ \partial \zeta$.
% (we have assumed that $\lambda$ does not depend on $j$). 
In the limiting case of sharp scattering potential, we find $s_1(\zeta) = \left[ J_0 + J_2\right]^2$ and
\bea
\mathcal{S}_1 (\zeta) &=&  \left[ J_0 + J_2\right] 
\left[ J_0 - 3 J_2 \right]\,, \nn \\
\mathcal{H}_1(\zeta) &=& J_0^2-6J_1^2+5J_2^2-2J_0J_2+2J_1J_3\,.
%31
\label{eq.hc}
\eea
If $\zeta \gg 1$, analytical results are obtained for an arbitrary correlator of impurity potential:
\begin{equation}
\mathcal{S}_1 (\zeta) \approx  - \frac{\ttr}{\tpi}  \frac{8 \cos 2 \zeta}{\pi \zeta^2}\,,~~ \mathcal{H}_1 (\zeta) 
\approx  \frac{\ttr}{\tpi} \frac{16 \sin 2 \zeta}{\pi \zeta}\,,
%32
\label{eq.hca1}
\end{equation}
where $\tpi$ differs from $\ttr$ by the substitution $w(q_{\varepsilon_F,\theta}) \rightarrow w(q_{\varepsilon_F,\pi})$ in \req{eq.tau1}.  
The function $\mathcal{H}_1(\zeta)$ and its high-field asymptote, given 
by \req{eq.hca1}, are shown in \rfig{fig1} by solid and dashed curves ``a'', respectively. 
Notice that for sharp scattering potential ($\ttr/\tpi=1$) \req{eq.hca1} is an excellent approximation already at $\epsilon \gtrsim 1$.
For comparison we also include curves ``b'' and ``c'', which represent different cases of smooth disorder, see figure caption.

%%%%%%%%%%%%%%%%%%%%%%%%%
%fig 1
%%%%%%%%%%%%%%%%%%%%%%%%%

The oscillations of $\mathcal{H}_1 (\zeta)$ describe HIRO for massless Dirac fermions, 
\req{eq.hiro1}, which differ from HIRO for conventional 2DEG, \req{eq.hiro},  
by a phase shift and by a decay $1/\edc$. These changes are the consequences of 
modification of the angular dependence of scattering probability and of its considerable 
reduction in the vicinity of backscattering ($\theta \simeq \pi$), owing to the factor 
$F^{(1)}_{\theta}$.
A similar phase shift and a stronger decay, for the same reasons, are also present 
in $\mathcal{S}_1 (\zeta)$ describing field effect on SdHO. It is unlikely, however, if 
$\mathcal{S}_1 (\zeta)$ can be accessed experimentally because heating of electrons by 
current, leading to a dependence of $T_e$ on $j$, becomes important already at 
$\zeta \ll 1$ and the nonlinearity associated with the first term in braces of 
\req{eq.r} prevails \citep{tan:2011}. 

In contrast to HIRO, the weak-field ($\zeta \ll 1$) nonlinear response of Dirac fermions, dominated 
(in the absence of SdHO) by $\mathcal{G}_1 (\zeta)$, is very similar to the one for conventional 2DEG. 
This response is rather insensitive to smoothness of disorder, but sensitive to the ratio $\ttr/\tin$. 
At $\edc > 1$, the function $\mathcal{G}_1 (\zeta)$ rapidly decreases as $\edc^{-3}$, much faster than $\epsilon^{-1}$ in conventional 2DEG.

It is worth noting that there exists another phenomenon which crucially depends on backscattering, 
the magnetophonon oscillations of linear resistance due to interaction of 2DEG with acoustic 
phonons (also known as the phonon-induced resistance oscillations, PIRO) \citep{zudov:2001b}. 
Recent observations of this phenomenon in graphene \citep{kumaravadivel:2019,greenaway:2019}
show that PIRO {\em do not} shift their phase and behave just like those in conventional 2DEG 
with parabolic band. This happens because the electron-phonon interaction in graphene is 
dominated by the gauge-field mechanism for which the interaction potential is not a scalar 
in the sublattice space and, as a result, backscattering is not suppressed. The microwave-induced 
resistance oscillations \citep{zudov:2001a}, which have not yet been observed in graphene, 
are not expected to change their phase either, as they are not sensitive to backscattering. 
Therefore, among the magneto-oscillatory phenomena specific for 2DEG in the regime of 
high LLs \citep{dmitriev:2012} only HIRO is expected to show profound changes in graphene, 
which makes them a special and promising tool for experimental probing of backscattering. 
   
In summary, we have developed a theory of nonlinear magnetoresistance for 2D fermions with Berry 
phase $\Phi_n=n \pi$, based on a model unifying the conventional 2DEG, massless Dirac fermions in 
graphene, and fermions in graphite bilayer. 
We have shown that the amplitude and the phase of nonlinear magnetoresistance oscillations of degenerate 2D fermion gas crucially depends on the parity of $n$. 
Such a distinction is a consequence of suppression of backscattering off impurity potential in the case of odd $n$. 
We believe that our results will simulate nonlinear magnetotransport experiments in graphene and in other topological materials.

\begin{acknowledgments}
We thank I. Dmitriev and M. Khodas for discussions.
The work at Minnesota was supported by the U.S. Department of Energy, Office of Science, Basic Energy Sciences, 
under Award \# ER 46640-SC0002567.
\end{acknowledgments}

%\bibliographystyle{../../apsrev-titles-t}
%\bibliography{../../bibRMP1qs.1_3,footnotes}

\end{document}